\documentclass[prb,aps,showpacs,twocolumn,unsortedaddress,superscriptaddress]{revtex4-1}
\usepackage{graphics, bm, psfrag, amsmath, amssymb, epsfig, grffile, float}
\usepackage{subfigure}
\usepackage{color}

\begin{document}

\title{Many-body instabilities and mass generation in slow Dirac materials}

\author{Christopher Triola}
\affiliation{Department of Physics, College of William and Mary, Williamsburg, Virginia 23187, USA}
\affiliation{Institute for Materials Science, Los Alamos National Laboratory, Los Alamos New Mexico 87545, USA}
\author{Jian-Xin Zhu}
\affiliation{Theoretical Division, Los Alamos National Laboratory, Los Alamos, New Mexico 87545, USA}
\affiliation{Center for Integrated Nanotechnology, Los Alamos National Laboratory, Los Alamos, New Mexico 87545, USA}
\author{Albert Migliori}
\affiliation{Glenn T. Seaborg Institute, Los Alamos National Laboratory, Los Alamos, New Mexico 87545, USA}
\author{Alexander V. Balatsky}
\affiliation{Institute for Materials Science, Los Alamos National Laboratory, Los Alamos New Mexico 87545, USA}
\affiliation{Nordic Institute for Theoretical Physics (NORDITA), Roslagstullsbacken 23, S-106 91 Stockholm, Sweden}


\begin{abstract}

Some Kondo insulators are expected to possess topologically protected surface states with linear Dirac spectrum, the topological Kondo insulators. Because the bulk states of these systems typically have heavy effective electron masses, the surface states may exhibit extraordinarily small Fermi velocities that could force the effective fine structure constant of the surface states into the strong coupling regime. Using a tight-binding model we study the many-body instabilities of these systems and identify regions of parameter space in which the system exhibits spin density wave, and charge density wave order.

\end{abstract}


\maketitle

\section{Introduction}

In many metals and semiconductors the behavior of the low energy electronic states can be understood in terms of free quasiparticles with quadratic energy dispersion in momentum $\textbf{p}^2/2m^*$, where $m^*$ is a system-dependent effective mass \cite{balatsky2014dirac}. However, there are a number of materials whose low energy electronic states are better described as massless Dirac fermions, including the superfluid phase of $^3$He, high-temperature \textit{d}-wave superconductors, graphene, and the surface states of topological insulators \cite{balatsky2014dirac,dahal2010charge,fu2007topological_prl,fu2007topological_prb}. In these Dirac materials the kinetic energy is proportional to the momentum $vp$, just like massless relativistic particles but with a speed $v$ that depends on the details of the system. For example, in graphene $v\approx 10^6$m/s$\approx c/300$.

The fact that quasiparticles obey the Dirac equation instead of the Schr$\ddot{\text{o}}$dinger equation can affect a variety of electronic properties, for example, the integer quantum Hall effect and localization \cite{neto2009electronic}. Another important way the Dirac nature of the quasiparticles manifests itself is in the effect of interactions. If the quasiparticles of a system obey the Schr$\ddot{\text{o}}$dinger equation, then the ratio of the average interparticle Coulomb energy to the average kinetic energy, $r_s=E_{C}/E_{K}$, is related to the density by $r_s\propto n^{-1/d}$, \cite{kotov2012electron,dahal2006absence} where the constant of proportionality depends on characteristics of the material. In contrast to normal metals, for Dirac materials this ratio is a characteristic of the system, independent of the electron density, given by $\alpha\equiv E_{C}/E_{K}=e^2/(\hbar\epsilon v)$. In this expression $e$ is the charge of the electron, $\epsilon$ is the material's dielectric constant, $\hbar$ is the reduced Planck constant, and $v$ is the speed of the Dirac particles. Much work has gone into the study of the phase diagram of graphene with respect to this parameter $\alpha$ \cite{dahal2010charge,dahal2006absence,kotov2012electron,drut2009graphene,khveshchenko2009massive,ryu2009masses,gorbar2002magnetic,gamayun2010gap,gamayun2009supercritical} and the results indicate that there is a critical value $\alpha_c$ such that if $\alpha<\alpha_c$ the spectrum remains gapless and if $\alpha>\alpha_c$ the system flows toward the strong coupling regime and is likely to develop a gap \cite{kotov2012electron}. Thus far, perturbative and numerical results suggest the critical value is $\alpha_c\approx 1$ \cite{kotov2012electron,drut2009graphene,gamayun2010gap,gamayun2009supercritical}, while experiments involving suspended graphene, for which $\alpha\approx 2.2$, seem to indicate a gapless state to within 0.1 meV of the Dirac point \cite{elias2011dirac}. Therefore, the ground state of Dirac materials in the strong coupling regime is not currently understood. For this reason we propose studying a class of materials with much smaller Fermi velocity than that of graphene since this class of materials is likely to possess $\alpha\gg\alpha_c$ and would be a better candidate for experiments probing the strong coupling regime in Dirac materials.

The surface of a three-dimensional (3D) topological insulator (TI) hosts two-dimensional (2D) Dirac quasiparticles similar to those found in graphene. Examples of experimentally verified 3D TIs include Bi$_2$Se$_3$, Bi$_2$Te$_3$, and Sb$_2$Te$_3$, all of which have Fermi velocities roughly half of that in graphene \cite{zhang2009topological,triola2014prl}. However, there is another class of topological insulators, the topological Kondo insulators (TKI), in which the bulk states are formed by renormalized $f$-electron levels which hybridize with conduction electrons to form a millivolt-scale gap in the bulk spectrum \cite{coleman2007heavy,dzero2012theory,dzero2010topological,roy2014surface}. The small gap in these materials combined with the large bulk effective mass imply that the surface Fermi velocity could be quite small. Some materials theoretically predicted to fall into this category include SmB$_6$ \cite{alexandrov2013cubic}, YbB$_{12}$ \cite{weng2014topological}, and PuB$_6$ \cite{deng2013plutonium}. Furthermore, there is a growing body of experimental evidence demonstrating that SmB$_6$ does in fact host metallic surface states \cite{neupane2013surface,jiang2013observation,kim2013surface,xu2013surface,wolgast2013low,xu2014direct,kim2014topological}.

Previous work has explored the possibility of broken symmetry states on the surface of a TKI employing a continuum model \cite{roy2014excitonic}. In this paper we present a tight-binding model to study the surface states of a TKI and proceed to investigate the possible ordered ground states for these systems within a mean field theory. From this analysis, we find regions of parameter space for the model that admit spin density wave and charge density wave solutions. For the case of strictly repulsive interactions we find that these ordered solutions lie within the region of parameter space corresponding to the strong coupling regime of Dirac materials ($\alpha > \alpha_c \approx 1$).

\section{Theoretical Model and Methods}
\begin{figure}[htb]
 \begin{center}
  \centering
  \includegraphics[width=9.0cm]{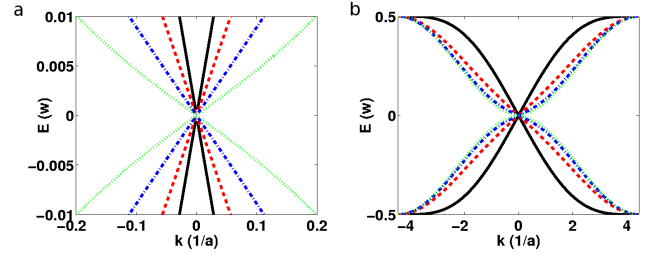}
  \caption{(Color online) Plots of the band structure given by Eq. (\ref{eq:bands_full}) along the diagonal of the square Brillouin zone using four different values of the parameter $A$ [$1/4$ (solid, black), $1/8$ (dashed, red), $1/16$ (dashed-dot, blue), and $1/32$ (dotted, green)], in two different momentum ranges [(a) from $\textbf{k}=(-1/5a,-1/5a)$ to $\textbf{k}=(1/5a,1/5a)$ and (b) from $\textbf{k}=(-\pi/a,-\pi/a)$ to $\textbf{k}=(\pi/a,\pi/a)$]. All energies are in units of the bandwidth.
         }
  \label{fig:bands}
 \end{center}
\end{figure}

To model the 2D surface states of a TKI, we consider a Hamiltonian defined on a square lattice:
\begin{equation}
\begin{aligned}
H_0 &= -i\dfrac{A}{2}\sum_{\alpha,\beta,\sigma}\sum_{\langle ij \rangle} \psi^\dagger_{i,\alpha,\sigma}\hat{z}\cdot\left( \hat{R}_{ij} \times \boldsymbol\sigma\right)_{\alpha\beta}\psi_{j,\beta,\sigma} \\
&+ \sum_{\alpha,\beta,\sigma}\sum_{i,j}\Gamma_{ij}\psi^\dagger_{i,\alpha,\sigma}\sigma_{\alpha\beta}\psi_{j,\beta,\sigma}
\end{aligned}
\label{eq:H0}
\end{equation}
where $\alpha$ and $\beta$ are orbital indices, $\sigma$ is a spin index, $\hat{R}_{ij}$ is the unit vector pointing from lattice site $j$ to lattice site $i$, and the matrix $\hat{\Gamma}$ is defined as
\begin{equation}
\Gamma_{ij} = \left\{\begin{array}{ll} 4\Gamma & ; \ i=j \\
 -\Gamma & ; \ i,j\text{ nearest neighbors} \\
 0 & ; \ \text{otherwise}   \end{array}\right. .
\label{eq:Gamma}
\end{equation}
The term proportional to $A$ leads to the formation of four separate Dirac points in the Brillouin zone. The term proportional to $\Gamma$ acts as a momentum-dependent mass term which gaps out all of the Dirac points except the one at $\textbf{k}=0$ allowing the model to represent the surface states of a strong TI \cite{vafek_2014}. The energy eigenvalues associated with this Hamiltonian in $k$ space are given by:
\begin{equation}
\begin{aligned}
E^{\pm}_{\textbf{k}}&=\pm4\Gamma\left[\sin^2\frac{ak_x}{2} + \sin^2\frac{ak_y}{2}\right] \\
&\times\sqrt{1+\left(\frac{A}{4\Gamma}\right)^2\frac{\sin^2ak_x + \sin^2ak_y}{\left[\sin^2\frac{ak_x}{2} + \sin^2\frac{ak_y}{2}\right]^2}}.
\end{aligned}
\label{eq:bands_full}
\end{equation}
Expanding this dispersion for small $k$ along the $k_x$ direction we find:
\begin{equation}
E^{\pm}_{k}\approx \pm \left(aAk + \dfrac{(3\Gamma^2-A^2)a^3}{6A}k^3\right).
\label{eq:bands_small-k}
\end{equation}
Thus, we can see that to first order in $k$ the dispersion matches the Dirac dispersion with Fermi velocity given by $aA/\hbar$. In Fig. \ref{fig:bands} we plot the full dispersion from Eq. (\ref{eq:bands_full}) for different ranges of $k$ to demonstrate the Dirac dispersion for a few different values of the Fermi velocity. It shows that near the Dirac point the parameter $A$ controls the Fermi velocity; however, for $A\ll\Gamma$ we can see that the cubic term in Eq. (\ref{eq:bands_small-k}) begins to dominate and the dispersion away from the Dirac point becomes noticeably less linear. Since we are most interested in the regime in which the model best describes a Dirac material, in this paper we focus on the case in which the chemical potential is close to the Dirac point.

In terms of the model parameters the bandwidth is given by:
$$w = \left\{\begin{array}{ll} 16\Gamma & ; \ A\leq4\Gamma \\
 \dfrac{4A^2}{\sqrt{2A^2+16\Gamma^2}} & ; \ A>4\Gamma   \end{array}\right.$$
Note that for $A\leq 4\Gamma$ the bandwidth is a constant set by the model parameter $\Gamma$. In the analysis that follows we restrict the range of $A$ to $A\leq 4\Gamma$ and present all energies in units of the bandwidth $w=16\Gamma$. We also present all distances in units of the lattice constant $a$.

To account for interactions we consider the full Hamiltonian: $H=H_0+H_I$, where $H_I$ takes on the exact form:
\begin{equation}
\begin{aligned}
H_I &= \dfrac{V_0}{2}\sum_{\sigma,\sigma'}\sum_{i\neq j} \dfrac{e^{-\mid \textbf{r}_i - \textbf{r}_j\mid/\lambda}}{\sqrt{\mid \textbf{r}_i - \textbf{r}_j\mid^2+d^2}} \psi^\dagger_{i,f,\sigma}\psi_{i,f,\sigma} \psi^\dagger_{j,f,\sigma'}\psi_{j,f,\sigma'} \\
&- U\sum_{i}  \psi^\dagger_{i,f,\uparrow}\psi_{i,f,\uparrow} \psi^\dagger_{i,f,\downarrow}\psi_{i,f,\downarrow}
\end{aligned}
\label{eq:interactions_full}
\end{equation}
where $V_0$ controls the strength of the long-range Coulomb interaction between $f$ electrons and $U$ is introduced as an on-site interaction between $f$ electrons.

In our calculations, we replace the exact interaction term $H_I$ with the mean field Hamiltonian:
\begin{equation}
\begin{aligned}
H^{MF}_I &= V_0\sum_{\sigma,\sigma'}\sum_{i\neq j} \dfrac{e^{-\mid \textbf{r}_i - \textbf{r}_j\mid/\lambda}}{\sqrt{\mid \textbf{r}_i - \textbf{r}_j\mid^2+d^2}} \langle n_{i,f,\sigma} \rangle \psi^\dagger_{j,f,\sigma'}\psi_{j,f,\sigma'} \\
&- U \sum_{i} \left( \langle n_{i,f,\uparrow} \rangle \psi^\dagger_{i,f,\downarrow}\psi_{i,f,\downarrow} + \langle n_{i,f,\downarrow} \rangle \psi^\dagger_{i,f,\uparrow}\psi_{i,f,\uparrow}  \right) \\
&+ \sum_{i} \left( \Delta_i \psi^\dagger_{i,f,\uparrow}\psi^\dagger_{j,f,\downarrow} + \Delta^*_i \psi_{i,f,\downarrow}\psi_{j,f,\uparrow }\right) + E_0.
\end{aligned}
\label{eq:interactions_mf}
\end{equation}
To capture the on-site Coulomb repulsion between $f$ electrons we can set $U=-V_0/d$. If we wish to include an attractive on-site interaction we set $U>0$. In the absence of a large electron-phonon coupling this attractive interaction could be engineered by the adsorption of a finite density of nonmagnetic molecules as discussed previously in the context of topological insulators \cite{she2014negative}. We may choose to write this in a more compact notation as:
\begin{align*}
H^{MF}_I &= \sum_{\alpha,\beta,\sigma,\sigma'}\sum_{i,j} W_{i\alpha\sigma,j\beta\sigma'} \psi^\dagger_{j,\beta,\sigma'}\psi_{j,\beta,\sigma'} \\
&+ \sum_{i} \left( \Delta_i \psi^\dagger_{i,f,\uparrow}\psi^\dagger_{j,f,\downarrow} + \Delta^*_i \psi_{i,f,\downarrow}\psi_{j,f,\uparrow }\right) + E_0
\end{align*}
where
$$W_{i\alpha\sigma,j\beta\sigma'}=\left\{\begin{array}{ll}  V_0\frac{e^{-\mid \textbf{r}_i - \textbf{r}_j\mid/\lambda}}{\sqrt{\mid \textbf{r}_i - \textbf{r}_j\mid^2+d^2}} \langle n_{i,f,\sigma} \rangle \delta_{\alpha\beta}\delta_{\alpha f}  & ; \ i\neq j \\
 -U\langle n_{i,f,\sigma} \rangle \delta_{\alpha\beta}\delta_{\alpha f} (1-\delta_{\sigma\sigma'}) & ; \ i=j   \end{array} \right. $$
and
$$\Delta_i \equiv U\langle \psi_{i,f,\uparrow}\psi_{j,f,\downarrow} \rangle.$$

Equipped with this mean-field Hamiltonian we perform a Bogoliubov transformation:
\begin{align*}
\psi_{i,\alpha,\uparrow} &=\sum_{n}\gamma_{n\uparrow}u_{i,\alpha,n,\uparrow} -\gamma^{\dagger}_{n\downarrow}v^*_{i,\alpha,n,\uparrow} \\
\psi_{i,\alpha,\downarrow} &=\sum_{n}\gamma_{n\downarrow}u_{i,\alpha,n,\downarrow} +\gamma^{\dagger}_{n\uparrow}v^*_{i,\alpha,n,\downarrow},
\end{align*}
where $\gamma^{\dagger}_{n\sigma}$ ($\gamma_{n\sigma}$) creates (annihilates) an eigenstate of the mean-field Hamiltonian $H$. It can be shown that the coefficients $u$ and $v$ satisfy the following equations:
\begin{equation}
\begin{aligned}
\epsilon_{n,\uparrow} u_{i,\alpha,n,\uparrow} &=\sum_{j,\beta}H_{i\alpha\uparrow,j\beta\uparrow} u_{j,\beta,n,\uparrow} + \Delta_i v_{i,f,n,\downarrow} \\
\epsilon_{n,\uparrow} v_{i,\alpha,n,\downarrow} &=-\sum_{j,\beta}H^*_{i\alpha\downarrow,j\beta\downarrow} v_{j,\beta,n,\downarrow} + \Delta^*_i u_{i,f,n,\uparrow} \\
\epsilon_{n,\downarrow} u_{i,\alpha,n,\downarrow} &=\sum_{j,\beta}H_{i\alpha\downarrow,j\beta\downarrow} u_{j,\beta,n,\downarrow} + \Delta_i v_{i,f,n,\uparrow} \\
\epsilon_{n,\downarrow} v_{i,\alpha,n,\uparrow} &=-\sum_{j,\beta}H^*_{i\alpha\uparrow,j\beta\uparrow} v_{j,\beta,n,\uparrow} + \Delta^*_i u_{i,f,n,\downarrow}
\end{aligned}
\label{eq:bdg}
\end{equation}
where $H_{i\alpha\sigma,j\beta\sigma'}\equiv H^{(0)}_{i\alpha\sigma,j\beta\sigma'} + W_{i\alpha\sigma,j\beta\sigma'}$ and $\epsilon_{n,\sigma}$ are eigenvalues of $H$.

Given the solutions to these equations we can write the mean fields as
\begin{equation}
\begin{aligned}
\langle n_{i,\alpha,\uparrow} \rangle &= \sum_{n} \mid u_{i,\alpha,n,\uparrow} \mid^2 f(\epsilon_{n,\uparrow}) \\
&+ \sum_{n} \mid v_{i,\alpha,n,\downarrow} \mid^2 \left(1-f(\epsilon_{n,\downarrow})\right) \\
\langle n_{i,\alpha,\downarrow} \rangle &= \sum_{n}  \mid u_{i,\alpha,n,\downarrow} \mid^2 f(\epsilon_{n,\downarrow}) \\
&+ \sum_{n} \mid v_{i,\alpha,n,\uparrow} \mid^2 \left(1-f(\epsilon_{n,\uparrow})\right) \\
\Delta_i &= U\sum_{n}  v^*_{i,f,n,\downarrow}u_{i,f,n,\uparrow}\left(1-f(\epsilon_{n,\uparrow})\right) \\
&- U\sum_{n} v^*_{i,f,n,\uparrow}u_{i,f,n,\downarrow}f(\epsilon_{n,\downarrow})
\end{aligned}
\label{eq:mean_fields}
\end{equation}
where $f(\epsilon)=\frac{1}{e^{\epsilon/k_BT}+1}$ is the Fermi-Dirac distribution function at temperature $T$ and $k_B$ is the Boltzmann constant. Given an initial set of model parameters and a temperature, Eqs. (\ref{eq:bdg}) and (\ref{eq:mean_fields}) allow us to solve for the density profile and superconducting order parameter $\Delta$ self-consistently. In the next section we discuss our progress toward solving these equations.

In some cases multiple solutions for the same model parameters may be found. In this case it is useful to compare the free energy associated with each of the solutions, given by $F=k_BT\ln Z$, where $Z$ is the partition function. The true ground state of the system will be given by the solution with the lowest free energy.

\section{Numerical Results and Discussion}

While it is straightforward to numerically solve Eqs. (\ref{eq:bdg}) and (\ref{eq:mean_fields}) for a finite system, we can make the computation more efficient by using the supercell technique\cite{jianxin_1999} as described in the appendix. For a system with a $10\times10$ real space unit cell and an $8\times8$ supercell we solved for self-consistent solutions to Eqs. (\ref{eq:bdg}) and (\ref{eq:mean_fields}). For the following results we focused on the case of nearest neighbor Coulomb interactions only and the zero temperature limit. In Eq (\ref{eq:interactions_mf}) we used a screening length of $\lambda=1$ and a lattice cutoff of $d=1$. We considered two limiting cases: the case of a repulsive on-site interaction ($U=-V_0$), and the case of an attractive on-site interaction that scales with the Coulomb interaction ($U=V_0$).

\begin{figure}[htb]
 \begin{center}
  \centering
  \includegraphics[width=8.0cm]{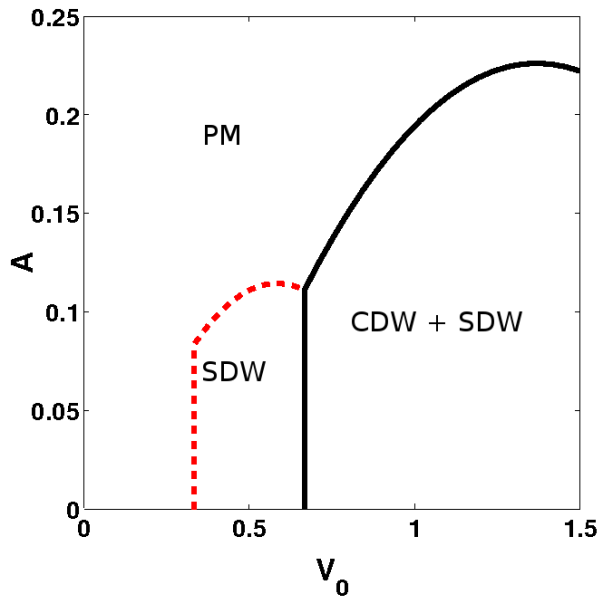}
  \caption{(Color online) Plot of the phases for the self-consistent solutions found in different regions of the $A$,$V_0$ plane. Note that the region
below the line $A= V_0$ appears to favor the formation of nontrivial order, consistent with $\alpha_c\approx 1$. The region enclosed by the red dashed line favors the formation of spin density wave order while in the region enclosed by the black solid line we find both spin density wave and charge density wave solutions. Outside of these regions the solution is paramagnetic (PM).}
  \label{fig:phase_1}
 \end{center}
\end{figure}

\begin{figure}[htb]
 \begin{center}
  \centering
  \includegraphics[width=8.0cm]{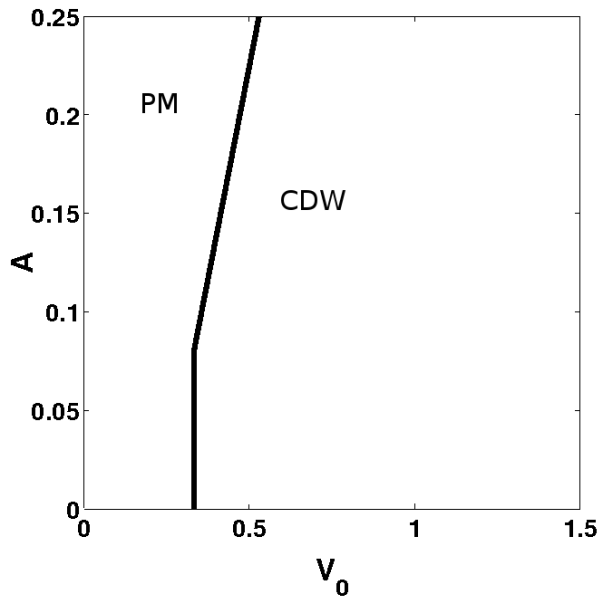}
  \caption{Plot of the phases for the self-consistent solutions found in different regions of the $A$,$V_0$ plane for attractive on-site interaction. In the region enclosed to the right of the solid black line the self-consistent solutions possessed charge density wave order; outside of this region the solution was paramagnetic (PM). }
  \label{fig:phase_2}
 \end{center}
\end{figure}

Starting from initial seeds that possessed antiferromagnetic, ferromagnetic, checkerboard and stripe charge density wave (CDW) order in addition to random seeds we found self-consistent solutions for Eqs. (\ref{eq:bdg}) and (\ref{eq:mean_fields}) using a convergence criterion of $10^{-3}$. Some of the self-consistent solutions that emerged from the different seeds for the same model parameters differed from each other. In these cases the one with the lowest free energy was taken to be the solution. In Fig. \ref{fig:phase_1} we show the regions of parameter space for which we found solutions in the case of repulsive on-site interactions while in Fig. \ref{fig:phase_2} we show the regions of parameter space for which we found solutions in the case with attractive on-site interactions.

First, we consider the case of on-site repulsion ($U=-V_0$), Fig. \ref{fig:phase_1}. Note that the general trend is consistent with our expectations for Dirac materials. In the region of strong coupling, $\alpha=V_0/A > \alpha_c$, we find Coulomb-driven ordered states, while in the weak coupling region, $V_0/A < \alpha_c$, a paramagnetic (PM) normal metallic state exists. These results are consistent with the established value of $\alpha_c\approx 1$. However, it appears that there is a critical value of the coupling, $V_c\approx w/3$, for this model below which the solution is trivial. This is in contrast to the case of a Dirac continuum model in which the only parameter governing the Coulomb interaction is $\alpha$. This difference can be attributed to the fact that for very small values of $A$ the band structure appears less linear and eventually the cubic term becomes more important, as we can see from Eq. (\ref{eq:bands_small-k}). It is reasonable to expect that real materials which host slow Dirac states will typically have similar behavior since the bands for these materials are expected to develop nonzero curvature away from the Dirac point \cite{dzero2012theory,dzero2010topological,neupane2013surface,xu2014direct}.

\begin{figure}[htb]
 \begin{center}
  \centering
  \includegraphics[width=4.0cm]{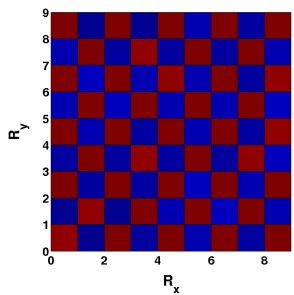}
  \caption{(Color online) Plot of the density modulations over a 10$\times$10 real space unit cell, as observed in the SDW and CDW regions shown in Figs. \ref{fig:phase_1} and \ref{fig:phase_2}.}
  \label{fig:cdw}
 \end{center}
\end{figure}

Taking a closer look at Fig. \ref{fig:phase_1} we can see that there are three distinct regions of the $V_0$,$A$ plane: a region favoring spin density wave (SDW) order, a region in which SDW and CDW coexist, and a region in which the solution was PM. Both SDW and CDW modulations were associated with $(\pi,\pi)$ wave vectors as shown in the sample plot in Fig. \ref{fig:cdw}. Intermediate states were also observed but these appear to be higher energy excitations. In the SDW region the boundary for the phase along the $V_0$ axis, at approximately one third of the bandwidth, defines the critical coupling, $V_c$. We find that above another critical value of $V_0$ CDW order begins to coexist with the SDW. In the coexistence region for some model parameters we were able to find solutions with exclusively CDW order but we lacked the resolution to see if these solutions indicated the existence of an additional region of the plane in which CDW is truly favored over SDW order, further calculations will be needed to answer this question.

Next we turn our attention to the case in which we include an on-site attraction ($U=V_0$), as shown in Fig. \ref{fig:phase_2}. In this case we find two regions: a region with CDW and a PM region. Again, these density modulations were associated with a wave vector of $(\pi,\pi)$. The region that favors CDW begins at $V_0\approx w/3$ and covers the rest of the plane. It is interesting to note that the CDW order appears for $V_0>w/3$ which is the same as $V_c$ for the case with repulsive on-site interactions. It should be noted that some of the self-consistent solutions we found near the transition region $V_0\approx w/3$ seemed to possess a small superconducting order parameter; however, this order parameter was usually just below the convergence criterion (even when the convergence criterion was lowered to $10^{-7}$). We attribute the absence of a superconducting region to the fact that we restricted ourselves to the case of half filling in which there was no density of states to allow for superconducting pairing. A more detailed study of the region near $V_0\approx w/3$ may be interesting for future work studying this model away from half filling.

Note in Fig. \ref{fig:phase_2} the absence of any regions with magnetic order, in contrast to Fig. \ref{fig:phase_1} in which both AFM and FM order were found. This can be accounted for by a heuristic argument based on Eq. (\ref{eq:interactions_mf}). Notice that the spin-dependent terms in the mean field Hamiltonian are given by $- U \sum_{i} \left( \langle n_{i,f,\uparrow} \rangle \psi^\dagger_{i,f,\downarrow}\psi_{i,f,\downarrow} + \langle n_{i,f,\downarrow} \rangle \psi^\dagger_{i,f,\uparrow}\psi_{i,f,\uparrow}  \right)$, thus the expectation value of the contribution to the total energy will be $- 2 U \sum_{i} \langle n_{i,f,\uparrow} \rangle \langle n_{i,f,\downarrow} \rangle$. For $U>0$ we can see that the energy can be minimized if the sum $\sum_{i} \langle n_{i,f,\uparrow} \rangle \langle n_{i,f,\downarrow} \rangle$ takes on its maximum possible value. Each term of this sum has a maximum value when $\langle n_{i,f,\uparrow} \rangle = \langle n_{i,f,\downarrow} \rangle=1/2$. Therefore the minimum energy can be expected to be achieved in a state with no magnetic order. However, for $U<0$ the minimum energy is achieved for a minimum value of $\sum_{i} \langle n_{i,f,\uparrow} \rangle \langle n_{i,f,\downarrow} \rangle$, which can allow the system to minimize its energy through an on-site spin polarization.

\section{Conclusions}

In conclusion, we presented a model for studying the surface states of a class of topological Kondo insulators and explored the dependence of the band structure on the model parameters, identifying the parameters which determine the Fermi velocity at the Dirac point. We then added interactions to this model, accounting for both Coulomb interactions as well as the possibility of an on-site attractive interaction. Using mean-field theory, at zero temperature, we found self-consistent solutions for different model parameters, investigating the relationship between the Fermi velocity at the Dirac point, the strength of the interactions, and the nature of the self-consistent solutions. For the case with on-site repulsion we identified three regions of parameter space with different Fermi velocity and coupling strength: a region which exclusively favored spin density wave order, a region of coexisting spin density wave and charge density wave order, and a paramagnetic normal metallic region. We also identified a critical value of the Coulomb interaction strength $V_c\approx w/3$ below which the solutions were normal metallic. When we considered the case of an attractive on-site interaction we found that the solutions possessed charge density wave order above this same critical Coulomb interaction strength.

Acknowledgements: We are grateful to Yuan-Yen Tai, Towfiq Ahmed, David Abergel, Yonatan Dubi, Yongkang Luo, Enrico Rossi, and Zachary Fisk for useful discussions. This work was supported by US DOE BES E304. The work of A.V.B. was also supported by ERC DM321031 and KAW. C.T. acknowledges the hospitality of the Seaborg Institute and the Institute for Materials Science at Los Alamos National Laboratory. This research used resources of the National Energy Research Scientific Computing Center (NERSC), a DOE Office of Science User Facility supported by the Office of Science of the U.S. Department of Energy under Contract No. DE-AC02-05CH11231.

\appendix

\section{Supercell Technique}

The system of equations given by Eqs. (\ref{eq:bdg}) and (\ref{eq:mean_fields}) can be solved for finite systems by simple matrix diagonalization. However, the matrix that must be diagonalized is $8N\times 8N$, where $N$ is the number of lattice sites and $8=2(\text{spins})\times 2(\text{orbitals})\times 2(\text{electron-hole})$. We can see that for 6400 sites this would involve diagonalizing a $51200\times 51200$ matrix which is not terribly practical. Using the supercell technique we can decrease the size of the matrix that needs to be diagonalized significantly. In the framework of the supercell technique we recognize that, due to the periodicity of the system, the solutions $u_{\textbf{r}_i,\alpha,n,\sigma}$ and $v_{\textbf{r}_i,\alpha,n,\sigma}$, are Bloch waves. To account for this we write
\begin{equation}
\begin{aligned}
u_{\textbf{r}_i,\alpha,n,\sigma}&= e^{i*\textbf{r}_i\cdot\textbf{k}}u_{\textbf{k},\textbf{r}_i,\alpha,n,\sigma} \\
v_{\textbf{r}_i,\alpha,n,\sigma}&= e^{i*\textbf{r}_i\cdot\textbf{k}}v_{\textbf{k},\textbf{r}_i,\alpha,n,\sigma},
\end{aligned}
\label{eq:bdg_bloch}
\end{equation}
where $\textbf{k}$ is the crystal momentum. After this transformation Eqs. (\ref{eq:bdg}) and (\ref{eq:mean_fields}) become:
\begin{equation}
\begin{aligned}
\epsilon_{\textbf{k},n,\uparrow} u_{\textbf{k},i,\alpha,n,\uparrow} &=\sum_{j,\beta}H_{i\alpha\uparrow,j\beta\uparrow;\textbf{k}} u_{\textbf{k},j,\beta,n,\uparrow} + \Delta_i v_{\textbf{k},i,f,n,\downarrow} \\
\epsilon_{\textbf{k},n,\uparrow} v_{\textbf{k},i,\alpha,n,\downarrow} &=-\sum_{j,\beta}H^*_{i\alpha\downarrow,j\beta\downarrow;\textbf{k}} v_{\textbf{k},j,\beta,n,\downarrow} + \Delta^*_i u_{\textbf{k},i,f,n,\uparrow} \\
\epsilon_{\textbf{k},n,\downarrow} u_{\textbf{k},i,\alpha,n,\downarrow} &=\sum_{j,\beta}H_{i\alpha\downarrow,j\beta\downarrow;\textbf{k}} u_{\textbf{k},j,\beta,n,\downarrow} + \Delta_i v_{\textbf{k},i,f,n,\uparrow} \\
\epsilon_{\textbf{k},n,\downarrow} v_{\textbf{k},i,\alpha,n,\uparrow} &=-\sum_{j,\beta}H^*_{i\alpha\uparrow,j\beta\uparrow;\textbf{k}} v_{\textbf{k},j,\beta,n,\uparrow} + \Delta^*_i u_{\textbf{k},i,f,n,\downarrow}
\end{aligned}
\label{eq:bdg_k}
\end{equation}
and
\begin{equation}
\begin{aligned}
\langle n_{i,\alpha,\uparrow} \rangle &= \dfrac{1}{M_{xy}}\sum_{n,\textbf{k}} \mid u_{\textbf{k},i,\alpha,n,\uparrow} \mid^2 f(\epsilon_{\textbf{k},n,\uparrow}) \\
&+ \dfrac{1}{M_{xy}}\sum_{n,\textbf{k}} \mid v_{\textbf{k},i,\alpha,n,\downarrow} \mid^2 \left(1-f(\epsilon_{\textbf{k},n,\downarrow}) \right)\\
\langle n_{i,\alpha,\downarrow} \rangle &= \dfrac{1}{M_{xy}}\sum_{n,\textbf{k}}  \mid u_{\textbf{k},i,\alpha,n,\downarrow} \mid^2 f(\epsilon_{\textbf{k},n,\downarrow}) \\
&+ \dfrac{1}{M_{xy}}\sum_{n,\textbf{k}} \mid v_{\textbf{k},i,\alpha,n,\uparrow} \mid^2 \left(1-f(\epsilon_{\textbf{k},n,\uparrow}) \right)\\
\Delta_i &= \dfrac{U}{M_{xy}}\sum_{n,\textbf{k}} v^*_{\textbf{k},i,f,n,\downarrow}u_{\textbf{k},i,f,n,\uparrow}\left(1-f(\epsilon_{\textbf{k},n,\uparrow})\right) \\
&-  \dfrac{U}{M_{xy}}\sum_{n,\textbf{k}} v^*_{\textbf{k},i,f,n,\uparrow}u_{\textbf{k},i,f,n,\downarrow}f(\epsilon_{\textbf{k},n,\downarrow}),
\end{aligned}
\label{eq:mean_fields_k}
\end{equation}
where $\textbf{k} = \frac{2\pi}{M_{xy} a}\left(\frac{n_x}{N_x},\frac{n_y}{N_y}\right)$ where $n_x=1,2,...,M_x$ and $n_y=1,2,...,M_y$, $M_x$ and $M_y$ are the number of unit cells in the $x$ and $y$ direction, respectively, $M_{xy}=M_xM_y$, $N_x$ and $N_y$ are the number of lattice sites per unit cell in the $x$ and $y$ direction, respectively, and we define
$$H_{i\alpha\sigma,j\beta\sigma';\textbf{k}}\equiv \sum_{\textbf{R}_j} e^{i\textbf{k}\cdot\left(\textbf{r}_j+\textbf{R}_j-\textbf{r}_i\right)}H_{\textbf{r}_i\alpha\sigma,(\textbf{r}_j+\textbf{R}_j)\beta\sigma'}.$$
Now, a system composed of 6400 sites can be studied by diagonalizing a $10\times10$ real space system using an $8\times8$ supercell. This means we only need to diagonalize a 800$\times$800 matrix instead of $51200\times 51200$. Moreover, this diagonalization is performed for each $\textbf{k}$ independently and thus the procedure may be easily parallelized to further improve performance.

\bibliographystyle{apsrev}
\bibliography{bib}

\end{document}